\documentclass[twocolumn,prb,amsmath,amssymb,floatfix,superscriptaddress,showpacs]{revtex4}
\usepackage{graphicx}
\usepackage{dcolumn}
\usepackage{bm}

\usepackage{times}
\begin{document}


\title{ The effect of local dipole moments on the structure and lattice dynamics of K$_{0.98}$Li$_{0.02}$TaO$_{3}$}
\author{Jinsheng Wen}
\affiliation{Condensed Matter Physics and Materials Science
Department, Brookhaven National Laboratory, Upton, New York 11973}
\affiliation{Department of Materials Science and Engineering, Stony Brook University,
Stony Brook, New York 11794}
\author{Guangyong Xu}
\affiliation{Condensed Matter Physics and Materials Science
Department, Brookhaven National Laboratory, Upton, New York 11973}
\author{C. Stock}\thanks{Also at ISIS, Rutherford Appleton Laboratory, UK.}
\affiliation{Physics Department, the Johns Hopkins University,
Baltimore, Maryland, 21218}
\affiliation{NIST Center for Neutron Research, National Institute of
Standards and Technology, Gaithersburg, Maryland 20899}
\author{P. M. Gehring}
\affiliation{NIST Center for Neutron Research, National Institute of
Standards and Technology, Gaithersburg, Maryland 20899}
\author{Z. Zhong}
\affiliation{National Synchrotron Light Source, Brookhaven National Laboratory,
Upton, New York 11973}
\author{L.A. Boatner}
\affiliation{Oak Ridge National Laboratory, Oak Ridge, Tennessee 37831}
\author{E. L. Venturini}
\author{G. A. Samara}
\affiliation{Sandia National Laboratories, Albuquerque, New Mexico
87185}
\date{\today}

\begin{abstract}
We present high energy x-ray (67\,keV) and neutron scattering
measurements on a single crystal of K$_{1-x}$Li$_x$TaO$_3$ for which
the Li content ($x=0.02$) is less than $x_c = 0.022$, the critical
value below which no structural phase transitions have been reported
in zero field. While the crystal lattice does remain cubic down to
$T=10$\,K under both zero-field and field-cooled ($E \le 4$\,kV/cm)
conditions, the Bragg peak intensity changes significantly at
$T_C=63$\,K. A strong and frequency-dependent dielectric
permittivity, a defining characteristic of relaxors is observed at
ambient pressure.  However an extensive search for static polar
nanoregions, which is also widely associated with relaxor materials,
detected no evidence of elastic neutron diffuse scattering between
300\,K and 10\,K. Neutron inelastic scattering methods were used to
characterize the transverse acoustic and optic phonons (TA and TO
modes) near the (200) and (002) Bragg peaks. The zone center TO mode
softens monotonically with cooling but never reaches zero energy in
either zero field or in external electric fields of up to 4\,kV/cm.
These results are consistent with the behavior expected for a
dipolar glass in which the local polar moments are frozen and
exhibit no long-range order at low temperatures.
\end{abstract}


\maketitle

\section{Introduction}

A defining feature of relaxor ferroelectrics is a large and highly
frequency-dependent dielectric permittivity that exhibits a broad
peak at a temperature that is not associated with a long-range
ordered structural phase transition.~\cite{GY1} Among such materials
the lead-oxide relaxors
$(1-x)$Pb(Mg$_{1/3}$Nb$_{2/3}$)O$_3$-$x$PbTiO$_3$
(PMN-$x$PT)~\cite{Xu_diffuse,Hiro_diffuse,PMN_diffuse3} and
$(1-x)$Pb(Zn$_{1/3}$Nb$_{2/3}$)O$_3$-$x$PbTiO$_3$
(PZN-$x$PT)~\cite{PZN_diffuse3,GXU3D} have attracted the greatest
attention because of the enormous potential they possess for use in
device applications and because of the interesting scientific
challenges they present to researchers attempting to understand the
physics of systems in which order and disorder coexist and compete.
Random fields arising from the heterovalent cations located on the
perovskite B-sites are believed to be a seminal ingredient that
underlies the properties of relaxors.~\cite{Stock1} At the same
time, KTaO$_3$, a perovskite compound that is known as an
"incipient" ferroelectric because it does not undergo a
ferroelectric transition in zero field~\cite{GY5} even though the
transverse optic (TO) phonon mode softens substantially at low
temperature~\cite{W2}, has long been of interest to scientists
because the material properties can be changed dramatically by
adding very small amounts of impurities.  The Li-doped material
K$_{1-x}$Li$_x$TaO$_3$, or KLT($x$), for example, transforms to a
tetragonal phase for Li concentrations as low as $x_c=0.022$ (below
this critical value no structural phase transition has been
observed).  Surprisingly, KLT($x$), which contains no heterovalent
cations and thus has comparatively little to no random fields, has
also been reported to exhibit dielectric properties characteristic
of relaxors for Li concentrations $x \ge x_c$.~\cite{GY4}

KLT($x$) has been extensively characterized using dielectric
spectroscopy, polarization hysteresis loops,~\cite{GY4,Toulouse_KLT}
Raman~\cite{W3,W5,Toulouse_KLT3}, and hyper-Raman scattering
techniques~\cite{W4}. As with other relaxor systems the concept of
polar nanoregions (PNR) has been invoked to explain much of the
experimental data on KLT($x$). These PNR are nanometer-scale regions
of randomly-oriented, local polarization that first appear at high
temperature within the paraelectric phase~\cite{Burns} and give rise
to strong, temperature-dependent diffuse scattering. However only a
few studies have published data on the x-ray~\cite{GY8} and neutron
diffuse scattering~\cite{Toulouse} observed in KLT($x$), which
represent the most direct evidence of PNR. For KLT(0.06) and
KLT(0.13) Yong {\it et al.}~\cite{KLT} found strong diffuse
scattering intensity around the (110) Bragg peak in the shape of
rods extending along cubic $\langle100\rangle$ directions, but no
diffuse intensity near (100) and (200).  Wakimoto {\it et
al.}~\cite{W} performed dielectric and neutron scattering
measurements on KLT(0.05) and also observed the formation of PNR
upon cooling to low temperature.  These studies suggest that the
behavior of PNR in KLT($x>0.022$) is similar to that reported in
other relaxors systems such as
PMN-$x$PT~\cite{Xu_diffuse,Hiro_diffuse,PMN_diffuse3} and
PZN-$x$PT~\cite{PZN_diffuse3,GXU3D}.  But the situation for KLT($x$)
at concentrations of Li below $x_c$ remains unclear.

\begin{figure}[ht]
\includegraphics[width=\linewidth]{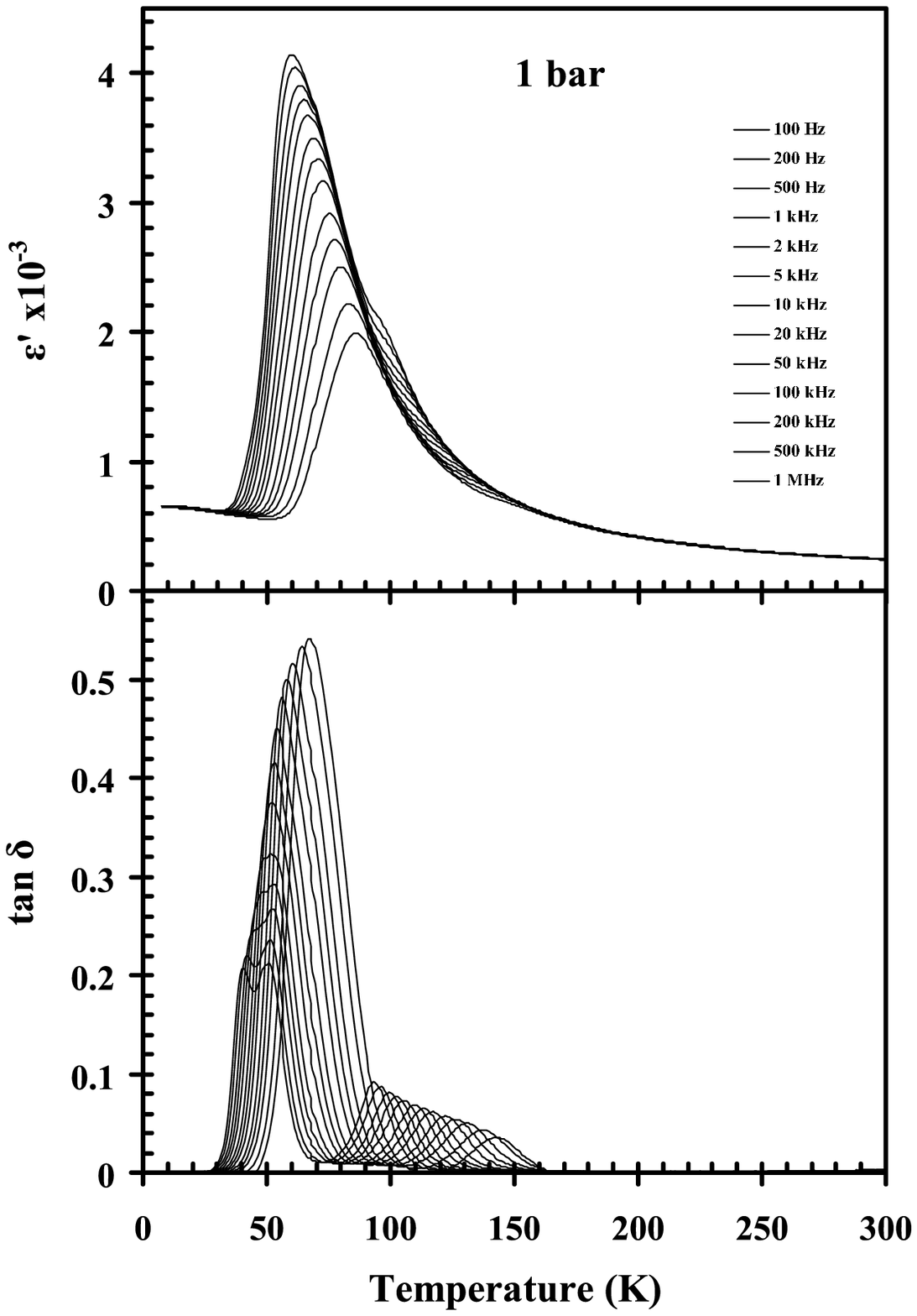}
\caption{Dielectric constant $\varepsilon^{\prime}$ and loss-tangent
tan$\delta$ of KLT(0.02) measured at different frequencies under a
pressure of 1 bar.} \label{fig:1}
\end{figure}

In this paper we focus on the properties of a KLT(0.02) single
crystal, for which the Li content lies just below the critical
concentration. Dielectric measurements show that the temperature
$T_m$, where the real part of the dielectric permittivity reaches a
maximum, shifts from about 60\,K to 90\,K as the measurement
frequency changes from 100\,Hz to 1\,MHz under a pressure of 1 bar
(Fig.~\ref{fig:1}). This behavior is believed to result from a
relaxation process involving local Li$^+$ dipole moments and
Li$^+$-Li$^+$ ion pairs. A key question is whether or not there are
local regions around these dipoles that are polar and form PNR. Both
high-energy x-ray and neutron scattering techniques have been used
to look for evidence of diffuse scattering that might indicate the
presence of PNR in KLT(0.02).  But in contrast to other relaxors and
KLT($x$) crystals with higher Li concentrations, no static diffuse
scattering intensity is observed in this sample down to $T=10$\,K.
This indicates that in KLT($x$) for $x<x_c$, the local Li moments
are mostly isolated and do not form larger (static) PNR.

Based on the shape of thermal diffuse scattering measured with
high-energy x-rays, we have confirmed that the TO phonon polarized
along $\langle100\rangle$ in this crystal is soft, which is
consistent with the (expected) tetragonal polarization scheme in KLT
compounds.  In addition, no explicit structural changes in the
crystal lattice structure are observed under both zero-field cooled
(ZFC) and field-cooled (FC) conditions. However there is an increase
in the Bragg peak intensity at $T_C=63$\,K, which suggests a
possible lowering of crystal symmetry through a release of
extinction. Low-energy acoustic and optic phonons were also studied
with neutron inelastic scattering methods. The ZFC phonon behavior
is very similar to that in the parent compound~\cite{KLTPHONON} and
in KLT(0.05)~\cite{W}, where a softening, but no condensation and/or
recovery, of the zone-center optic mode occurs.  When cooled under
an external electric field $E=4$\,kV/cm oriented along the [001]
direction, the crystal structure and energies of the TO/TA phonons
are barely affected, whereas the intensities  of long-wavelength
phonons show interesting changes. Our results suggest that for small
Li contents, KLT($x$) changes from an incipient ferroelectric to a
dipolar glass, where the dipole moments are frozen locally without
long-range ferroelectric order.

\section{Experiment}

The KLT(0.02) single crystal examined in this study has dimensions
0.5\,cm$\times$1\,cm$\times$2\,cm and was grown at Oak Ridge
National Laboratory. The crystal structure is cubic with a room
temperature lattice constant $a=3.992$~\AA. The Li concentration of
the crystal was estimated from the amount of Li in the melt and then
corroborated from the established relationship between the Li
concentration and the peak temperature of the dielectric
permittivity, investigated by dielectric spectroscopy with
measurements of the real ($\varepsilon^{\prime}$) and loss
(tan$\delta$) parts of the dielectric function.~\cite{W12} The set
up for dielectric measurements is the same as that described in
Ref.~\onlinecite{W}.

Neutron scattering experiments were carried out on the cold neutron
triple-axis spectrometer SPINS and the thermal neutron triple-axis
spectrometer BT7, which are located at the NIST Center for Neutron
Research (NCNR). Horizontal neutron beam collimations of
guide-80'-S-80'-open (S=sample) and 50'-50'-S-40'-240' were used for
the measurements on SPINS and BT7, respectively.  All data were
taken in a fixed final energy mode (5.0\,meV for SPINS and 14.7\,meV
for BT7) using the (002) Bragg reflection from highly-oriented
pyrolytic graphite (HOPG) crystals to monochromate the incident and
scattered neutrons.  During the experiments on SPINS a Be filter was
placed before and after the sample to reduce the scattering from
higher order reflections; a single HOPG filter was placed after the
sample on BT7 for the neutron inelastic measurements.  All data were
taken in the (H0L) scattering plane defined by the vectors [100] and
[001], and described in terms of reciprocal lattice unit (rlu),
where 1~rlu~$=a^*=2\pi/a=1.574$~\AA{}$^{-1}$. The electric field was
applied along the [001] direction during the FC measurements. All
data were taken on cooling from 300\,K to ensure that all residual
(poling) effects were removed.

X-ray scattering experiments were performed at beamlines X17B1 and
X22A located at the National Synchrotron Light Source (NSLS). A
67\,keV x-ray beam with an energy-resolution $\Delta E/E=10^{-4}$
was produced at X17B1 using a sagittal-focusing double-crystal Si
(311) monochromator with both crystals oriented in the asymmetric
Laue mode.~\cite{zhong01_1} Charge coupled device (CCD) detector
were used to perform monochromatic Laue-style measurements. In this
type of configuration scattering intensities on a large part of the
surface of the Ewald sphere can be measured simultaneously.  Data
were taken in the (HK0) and (HKK) zones. Further details about the
x-ray diffuse scattering experimental set up are discussed in the
paper by Xu {\it et al.}.~\cite{GXU3D} Structural measurements were
also performed on the X22A beamline using an incident x-ray energy
of 10.7\,keV and a perfect Si crystal analyzer.

\section{Results and Discussion}

\subsection{ Structure }

\begin{figure}[ht]
\includegraphics[height=\linewidth,angle=90]{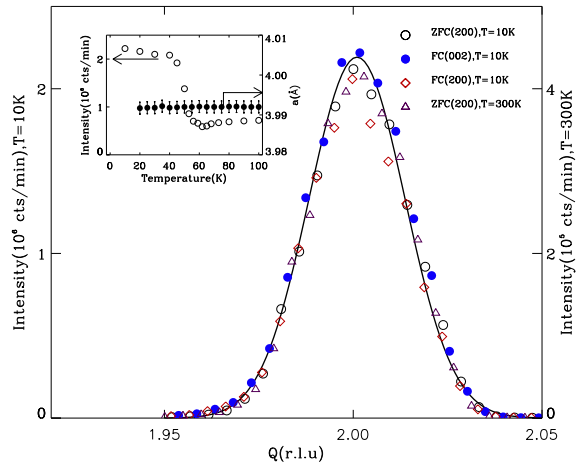}
\caption{(Color online) ZFC (200) Bragg peak scan at 300\,K, ZFC and
FC (200) and (002) Bragg peak scans at 10\,K. The intensities of the
300\,K data were scaled to permit a direct comparison with those at
10\,K. The inset shows the ZFC (200) Bragg peak intensity measured
with neutrons on BT-7 and the ZFC lattice parameters measured with
x-rays on X22A. Uncertainties in the Bragg intensities are
commensurate with the scatter in the data.  The error bars in the
inset are obtained by least-square fitting the data with Gaussian
functions. The solid line is a guide for the eyes.} \label{fig:2}
\end{figure}

The structure of KLT(0.02) was examined using elastic neutron
scattering and x-ray diffraction methods. As shown in
Fig.~\ref{fig:2}, there is no discernable change in the position or
longitudinal width of the (200) Bragg peak between 300\,K and 10\,K.
There is also no obvious difference between Bragg peaks measured at
(200) and (002). Based on the BT7 instrumental wave-vector
resolution, which is $\Delta Q/Q \sim 5\times10^{-3}$ full width at
half maximum (FWHM), we can set an upper limit of $\alt0.1\%$ on the
tetragonality $c/a - 1$. Within this limit KLT(0.02) does not
undergo a tetragonal lattice distortion down to $T=10$\,K. This is
confirmed using higher resolution $\Delta Q/Q \sim 3\times10^{-4}$
x-ray diffraction measurements, with absence of change in lattice
constant at the entire temperature range. However, the ZFC (200)
Bragg peak intensity jumps at $T_C=63$\,K, as shown in the inset of
Fig.~\ref{fig:2}. The Bragg peak intensity change is observed
repeatedly under cooling and heating. This effect is quite common in
ferroelectric systems where a release of extinction occurs when the
system transforms to a lower symmetry phase, and a crystal breaks up
into many domains.~\cite{Stock1} In KLT(0.02), however, there is no
explicit tetragonal phase. In this case the increase in the Bragg
peak intensity at $T_C=63$\,K may be due to crystal imperfections,
which can result either from subtle shifts in the atomic positions
in an otherwise cubic lattice, or from local strain/stress induced
at low temperature.

The same measurements were performed on the sample after it was
cooled in an external electric field $E=4$\,kV/cm applied along
[001]. Neither the width nor the intensity of the Bragg peaks were
affected, and a same Bragg peak intensity increase occurs at 63~K.
Thus the application of a moderate electric field along [001] has no
discernable effect on the average static structure of this system.

\subsection{ Diffuse scattering }

\begin{figure}[ht]
\includegraphics[width=\linewidth]{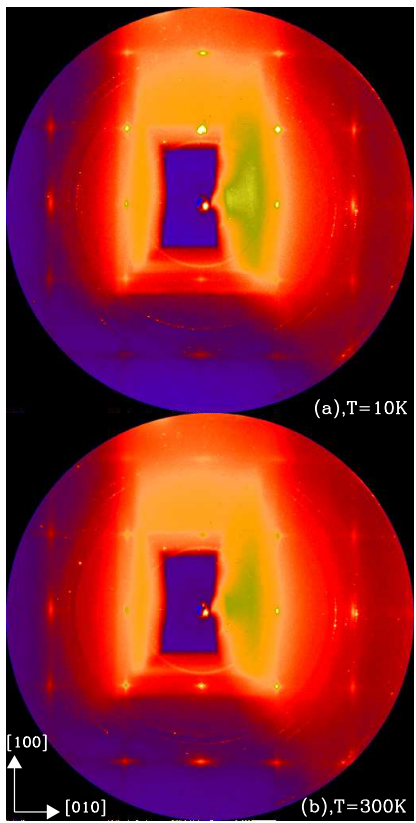}
\caption{(Color online) CCD images showing the diffuse scattering
from KLT(0.02) measured at $T=10$\,K and 300\,K in the (HK0) zone.
The incident x-ray beam is oriented along [001].} \label{diffuse1}
\end{figure}

Fig.~\ref{diffuse1} shows CCD images taken at 10\,K and 300\,K in
the (HK0) zone. In this plane the diffuse scattering extends along
the [100] or [010] direction, or both.  For example, near the
($\bar{2}$00) Bragg peak the diffuse scattering is transverse in
character in that it extends primarily along [010], whereas the
diffuse scattering near ($\bar{2}$10) extends along both [100] and
[010].  These observations hold true at 10\,K and 300\,K and are
qualitatively similar to those measured in other KLT($x$)
systems.~\cite{KLT,W} However the diffuse intensities increase with
temperature indicating that these measurements may be dominated by
thermal diffuse scattering.

\begin{figure}[ht]
\includegraphics[width=\linewidth]{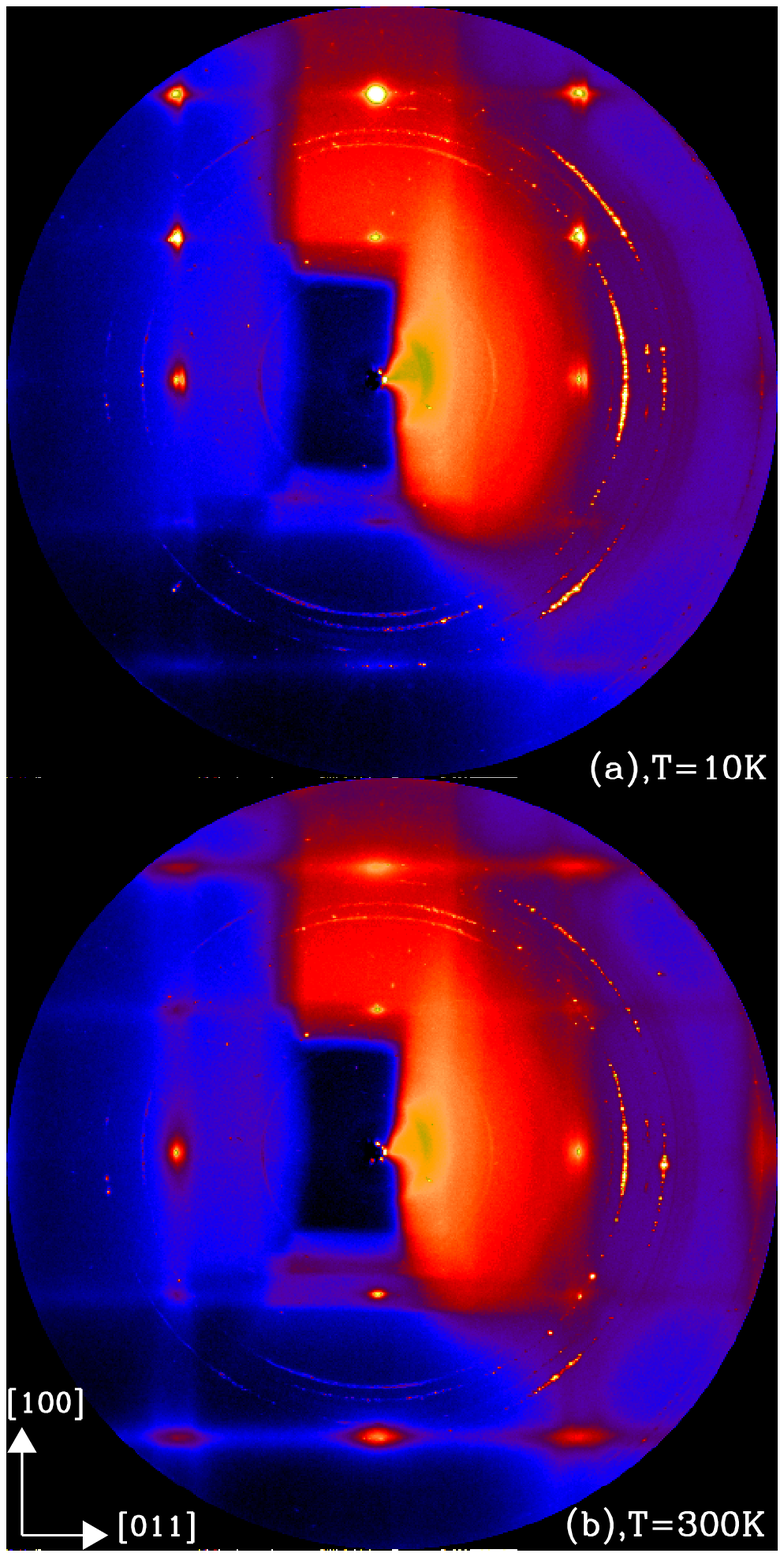}
\caption{(Color online) CCD images showing the diffuse scattering
from KLT(0.02) measured at 10\,K and 300\,K in the (HKK) zone. The
incident x-ray beam is oriented along [01$\bar{1}$].} \label{fig:3}
\end{figure}

To obtain a complete picture of the diffuse scattering in three
dimensions we rotated the sample about [100] by $45^\circ$ and
measured diffuse scattering intensities in the (HKK) zone.
Fig.~\ref{fig:3} shows the corresponding CCD images obtained at
10\,K and 300\,K.  The sample was also tilted about the [011]
direction by $2^\circ$ to better observe the out-of-plane components
of the diffuse scattering (for a detailed description of this
technique see Ref.~\onlinecite{GXU3D}). Near ($\bar{2}$00) we
observe diffuse scattering that extends along [011], whereas near
($\bar{1}$11) the diffuse scattering extends along both [100] and
[011]. Combining these data with those from the (HK0) zone, we find
that the diffuse scattering in KLT(0.02) forms \{001\} planes rather
than rods oriented along $\langle001\rangle$.  By contrast, the
diffuse scattering measured in the lead-oxide relaxor PZN forms
ellipsoids that are extended along $\langle110\rangle$.~\cite{GXU3D}
The planar geometry of the diffuse scattering in KLT(0.02) was
confirmed after we tilted the sample so that the Ewald sphere was
displaced even further from the Bragg peaks; then we could actually
observe a splitting of the diffuse scattering intensities in the
(010) and (001) plane, showing two almost vertical lines near (211)
on the CCD image. Note that the diffuse scattering from all three
planes ((100), (010), and (001)) is not always present around all
Bragg peaks. For example, near (200) only the intensity from the
(100) plane is observed while intensities in the (010) and (001)
planes are absent. This is because the neutron (and x-ray) diffuse
scattering cross section resulting from local atomic shifts is
proportional to $|{\bf Q}\cdot\epsilon|^2$, where $\epsilon$ is the
polarization vector.  We can therefore carry out a simple
polarization analysis similar to that done in
Ref.~\onlinecite{GXU3D}. Our analysis shows that the diffuse
scattering intensity from \{001\} planes arises from polarizations
oriented along $\langle001\rangle$, i.\ e.\ perpendicular to the
planes.  This indicates that the phonon mode polarized along
$\langle100\rangle$ is soft, which is perfectly natural for a system
that has a tetragonal ground state.  By comparison, in the case of
the lead-oxide relaxors (e.\ g.\ PMN-$x$PT and PZN-$x$PT) the phonon
mode polarized along $\langle110\rangle$ is soft and the ground
state is rhombohedral.~\cite{GXU3D,stock04}

One disadvantage of x-ray diffuse scattering measurements is the
lack of energy resolution; this makes it difficult to distinguish
dynamic contributions, e.\ g.\ phonons or dynamic PNR, from static
contributions.  There have been some reports~\cite{KLT,W} that the
diffuse scattering in KLT($x$) for $x>x_c$ is mainly static, as is
the case in the lead-oxide relaxors PMN-$x$PT~\cite{Xu_diffuse} and
PZN-$x$PT~\cite{GXU3D}. However, in KLT(0.02) the x-ray diffuse
scattering intensity increases with temperature, which is
inconsistent with the behavior observed in these other relaxor
systems where the diffuse scattering from static PNR decreases with
temperature.  Com\`{e}s and Shirane reported that phonon
contributions dominate the x-ray diffuse intensity in the parent
compound KTaO$_3$~\cite{KTaO3}, which strongly suggests that thermal
diffuse scattering may dominate in KLT($x$) samples with very low Li
contents. In order to resolve this issue for KLT(0.02), we performed
neutron scattering measurements using the NCNR SPINS spectrometer,
which provided very good energy resolution ($\sim 0.34$\,meV FWHM).

\begin{figure}[ht]
\includegraphics[width=\linewidth]{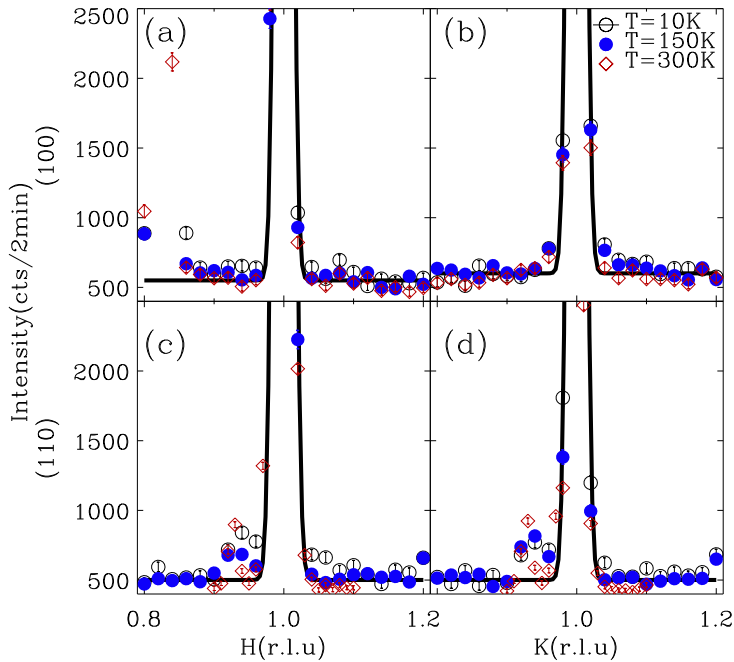}
\caption{(Color online) Elastic scans at 10\,K, 150\,K and 300\,K.
(a), (b), H and K scans at (100); (c), (d), H and K scans at (110).
Error bars represent square root of the counts. Lines are guides for
the eyes.} \label{fig:4}
\end{figure}

Fig.~\ref{fig:4} shows scans of the elastic scattering intensity as
a function of the wavevector measured on SPINS along [100] and [010]
at 10\,K, 150\,K and 300\,K near the (100) and (110) Bragg peaks.
The Bragg peaks were fit using resolution-limited Gaussian
functions. The background is $\sim 500$\,counts/2\,min and is nearly
temperature independent and identical at each Bragg peak.  Diffuse
scattering should be weak and broad compared to that from the Bragg
peaks, appearing as broad tails on either side of the Bragg peak.
However, for KLT(0.02) we find that outside of both Bragg peaks the
scattering intensity is flat and temperature independent, indicating
the absence of any elastic diffuse scattering.

Here we only measured diffuse scattering near (100) and (110) peaks.
Since it is highly unlikely that diffuse scattering structure
factors are weak near both Bragg peaks, we believe that the static
diffuse scattering is extremely weak in KLT(0.02). These results can
be compared directly to those on PMN measured on the same instrument
near these same two peaks.~\cite{Hiro_diffuse} The diffuse
scattering from KLT(0.02) is at least one to two orders of magnitude
weaker. We thus conclude that the diffuse x-ray scattering shown in
Fig.~\ref{diffuse1} and Fig.~\ref{fig:3} is thermal diffuse in
nature, i.\ e.\ dominated by phonons. This situation is quite
different from that observed in other relaxor systems, as the
diffuse scattering has been shown to be mainly elastic in
PMN-$x$PT~\cite{Xu_diffuse,matsuura:144107}, PZN-$x$PT~\cite{GXU3D},
and even KLT($x$) at higer Li concentrations~\cite{W,KLT}.  We also
measured the elastic diffuse scattering under an external electric
field oriented along [001].  Again in contrast to the behavior
observed in both PMN-$x$PT and PZN-$x$PT, no change in the diffuse
scattering was observed up to $E=4$\,kV/cm.

In KLT($x$) it is known that local Li displacements lead to local
dipole moments.  Yet despite the absence of any measurable diffuse
scattering in our sample, and thus by inference any static PNR,
these local Li dipoles apparently still contribute to the relaxation
process that leads to a strongly frequency-dependent dielectric
permittivity. The clear difference between KLT(0.02) and other
relaxor compounds in which PNR are observed~\cite{W, PMN_neutron,
PMN_neutron2, PMN_diffuse, Xu_diffuse, GXU3D} is that for KLT(0.02)
the local moments are indeed really ``local,'' i.\ e.\ they do not
polarize the surrounding regions to form PNR that exceed a few unit
cells in size.  Further, no correlations between local Li moments
are evident from our measurements. With increasing Li concentration,
the effect of these local polar moments on the bulk system becomes
stronger, as evidenced by both the existence of static diffuse
scattering (larger PNR) and an explicit phase transition into a low
temperature tetragonal phase (for $x>x_c$). Whether these two
effects are independent or correlated has yet to be determined.

\subsection{ Phonons under electric field }

\begin{figure}[ht]
\includegraphics[width=\linewidth]{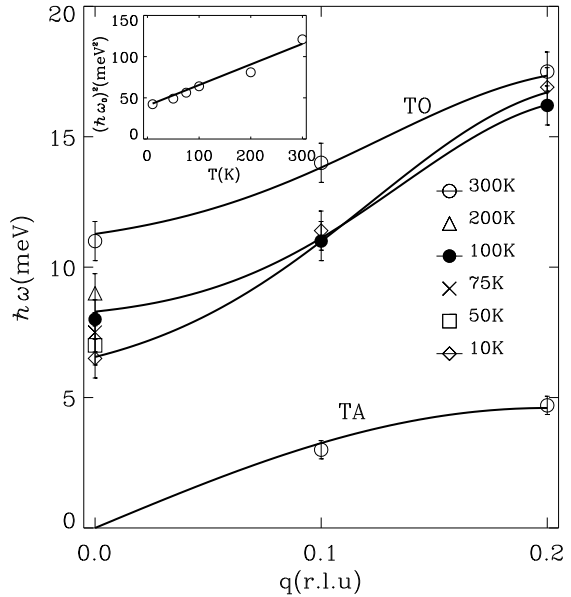}
\caption{Dispersion of TO and TA phonons under zero field at
different temperatures. Measurements were taken near (200) along
[0$\bar1$0]. The inset is a plot of $(\hbar\omega_0)^2$ vs.
temperature. Error bars are obtained by least-square fitting the
data with Lorentzian functions. Lines are guides for the eyes.}
\label{phonon2}
\end{figure}

The lowest-energy TO and transverse acoustic (TA) phonons have been
characterized near (200) and (002) using the BT7 thermal neutron
triple-axis spectrometer. In Fig.~\ref{phonon2} we plot the
dispersions for the TA and TO phonons measured near (200) along
[0$\bar1$0] at various temperatures under ZFC conditions. From
300\,K to 10\,K the zone-center TO phonon clearly softens, but it
never reaches zero energy. This behavior is very similar to that
seen in both the parent compound KaTO$_3$~\cite{KLTPHONON} and in
KLT(0.05)~\cite{W}. In fact at 10\,K the zone-center TO energy is
more than 6\,meV, which is higher than that ($\sim$3~meV) in the
parent compound KTaO$_3$.~\cite{KLTPHONON} Raman
measurements~\cite{W3} on a KLT($x$) crystal with a similar Li
content are consistent with ours for this TO phonon mode. These
results clarify that the x-ray diffuse scattering intensity
discussed previously arises from the TA phonon because the TA phonon
lies lower in energy than the TO phonon, which never drops below
6~meV.

If we plot the soft mode energy $(\hbar\omega_0)^2$ {\it v.s.}
temperature (see the inset to Fig.~\ref{phonon2}) we see that
$(\hbar\omega_0)^2$ decreases linearly with T as expected for a
conventional displacive ferroelectric.  This behavior is also
observed in the lead-oxide relaxors such as PMN at high temperature.
However, unlike the case of PMN, the soft mode in KLT(0.02) never
recovers as the linear decrease in $(\hbar\omega_0)^2$ persists down
to at least 10\,K. This is consistent with the absence of a
ferroelectric phase transition for $T>0$ and suggests that the
local, polar moments freeze on cooling, which implies the presence
of a low-temperature dipolar glass phase.  At the same time, the
energy width of the soft mode remains almost constant, yet larger
than the instrument resolution, thus indicating a short phonon
lifetime. This is similar to what is observed in KLT(0.05).~\cite{W}
It is thus possible that the local dipole moments resulting from the
large Li ionic displacements do interact with, or scatter, the soft
mode, even though there is no clear temperature scale associated
with this interaction.  In this respect the soft mode in KLT(0.02)
behaves quite differently from those measured in PMN-$x$PT, and
PZN-$x$PT, where the so-called "waterfall" effect (i.\ e.\ a
wavevector-dependent broadening of the soft mode in energy) is
observed~\cite{PMN_softmode, PZN_waterfall1, PZN_waterfall2} and
lately interpreted in terms of a defect model arising from chemical
and valence disorder~\cite{PMN-60PT}. In our KLT(0.02) sample the Li
content is relatively low; thus effects from chemical disorder are
minimal and no "waterfall" effect is present.

Interestingly, the TA mode in KLT(0.02) behaves quite differently
from that in KLT(0.05).~\cite{W}  Unlike the more highly Li-doped
samples, the TA phonon in our sample remains well-defined over the
temperature range (10\,K to 300\,K). There is virtually no
temperature dependent change in either the dispersion, lineshape, or
linewidth of the TA mode. This suggests that the broadening of the
TA mode reported in KLT(0.05) on cooling is very likely due to an
interaction between the the TA mode and the polar, short-range order
(PNR). It seems plausible that as the PNR in KLT(0.05) grow larger
on cooling, their interaction with the TA mode becomes stronger,
causing a change of the TA phonon line width (lifetime). In
KLT(0.02), since there is no static diffuse scattering, but only
isolated local Li moments, the TA phonon mode remains unchanged.

\begin{figure}[ht]
\includegraphics[width=\linewidth]{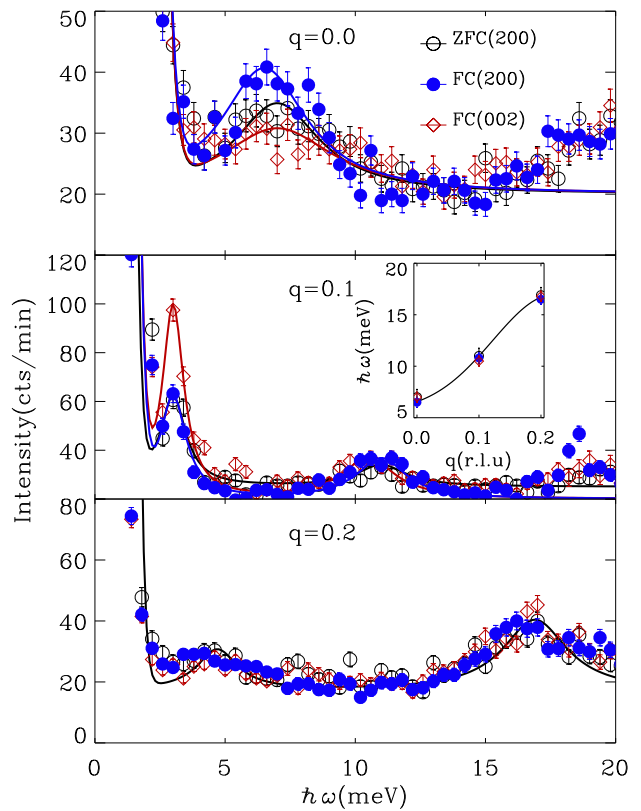}
\caption{(Color online) Constant $q$ scans for $q=0.0$, 0.1,
0.2~rlu, measured at 10~K. The inset shows TO phonon dispersion
measurements under ZFC (open circles), FC(200) (closed circles), and
FC(002) (diamonds) conditions. Intensity errors represent the square
root of the counts, and error bars in the inset are obtained by
least-square fitting the data with Lorentzian functions. Lines are
guides for the eyes.} \label{phonon3}
\end{figure}

When an electric field is applied along the [001] direction, we
observe interesting changes in the phonon behavior. As shown by our
structural measurements, no long-range, ferroelectric, tetragonal
phase is induced when cooling under a moderate field. The phonon
behavior is completely consistent with our structural results. At
10\,K, the energy of the TA mode is unaffected by the external
field, and any effect on the energies of the TO mode is very small
(see the inset to Fig.~\ref{phonon3}).  In other words, the FC
energies of the TA and TO phonons remain close to the ZFC values in
our sample in which there is no explicit low-temperature tetragonal
phase. This finding stands in stark contrast to those measured in
more highly doped KLT($x$) samples, e.\ g.\ in
KLT(0.035)~\cite{Toulouse} where a clear splitting of the TA mode is
observed when the system is cooled under a [001] field and
transforms into a tetragonal phase.  Ostensibly this is because the
local dipole moments in KLT(0.02) are frozen, thus making it very
difficult to drive the system into a long-range-ordered
ferroelectric phase with an external field.

Despite the absence of a measurable change in phonon energies under
an external field there is a significant change in the intensities
and/or lineshapes of the TA/TO phonons measured near the (200) and
(002) Bragg peaks.  The intensity for the zone center TO mode
measured near (200) increases and the peak becomes sharper, while
near (002) the intensity of the TO mode decreases, and the peak
becomes broader in energy. The difference between the intensities is
as large as 50\%. As $q$ increases the effect appears to diminish.
When $q=0.1$~rlu, the difference in the optic modes measured near
(200) and (002) is already tiny.  At a wavevector of $q=0.2$~rlu,
the TA and TO phonons measured near (200) and (002) are very
similar. We are puzzled, however, to note that the FC acoustic mode
intensity near (002) increases significantly.

These changes suggest that even though no clear evidence of any
long-range polar order is induced with the application of a moderate
electric field oriented along [001], a field can still influence the
overall polar structure of the system to a certain extent. It is
likely that an electric field can lead to an enhancement of the
local Li dipole moments polarized along [001]. This will lead to a
stronger interaction between these local dipole moments and TO
phonons having the same polarization ([001]). The result is the
broadening of the zone center TO mode measured near (002). The
change in the TA phonon mode indicates a strong coupling between TA
and TO modes in the system, similar to that in its parent system
KTaO$_3$~\cite{KTaO3}, which should only be significant for small
(but non-zero) $q$.

\section{Summary}

The strong frequency-dependent peak in the dielectric curve
$\varepsilon^\prime(\omega,T)$ observed in KLT(0.02) would appear to
justify categorizing it with other relaxor systems. Yet unlike other
relaxors where static PNR are easily identified by the presence of
strong, temperature dependent diffuse scattering, no such scattering
is observed near either the (100) and (110) Bragg peaks between
10\,K and 300\,K.  The diffuse planar diffuse scattering intensities
observed with x-rays are dominated by contributions from soft phonon
modes. This makes the system very unique. In KLT(0.02) the PNR are
either small in volume (i.\ e. isolated local Li dipole moments) or
dynamic in nature. In fact, these local Li dipole moments dominate
the physics in this system.

Our structural measurements show no splitting of the main (200)
Bragg peak in either ZFC or FC conditions. This suggests that the
shape of the unit cell in KLT(0.02) remains cubic down to low
temperatures within the precision of our measurements ($\alt \sim
0.1\%$). The change in Bragg peak intensity at $T_C=63$\,K suggests
that subtle changes in the crystal do occur at $T_C$.  On the other
hand, although the zone center energy of the TO phonon softens
significantly with cooling, it does not condense at low temperature.
No ferroelectric phase transition takes place under ZFC or FC
conditions (for field strength up to 4~kV/cm), which is suggestive
of a dipole glass phase in which the local moments are frozen and
long-range polar order is not achieved.

Interesting electric field effects have been observed in our neutron
measurements of the TO and TA modes for small $q<0.20$~rlu. Although
electric fields of up to 4\,kV/cm oriented along [001] fail to
induce long-range polar order, the same fields appear to be able to
affect the Li moments such that long wavelength TO phonons polarized
along [001] are more strongly scattered by the local Li moments,
thus causing them to become broader in energy (shorter lifetimes).
However, the system remains in a dipolar glass state where the local
Li moments are frozen and no long-range correlation between these
moments exists, as also evidenced by the absence of static diffuse
scattering under an external field.

\begin{acknowledgements}
We would like to thank stimulating discussions with H.~J.~Kang,
J.~H.~Chung, J.~W.~Lynn, and Y.~Chen. The work at Brookhaven
National Laboratory was supported by the U.\ S.\ Department of
Energy (DOE) under contract No.~DE-AC02-98CH10886. C.\ Stock was
supported by Natural Sciences and Engineering Research Council of
Canada and the NSF under Grants No.\ DMR-0306940. Research at ORNL
is sponsored by the Division of Materials  Sciences and Engineering,
Office of Basic Energy Sciences, U.\ S.\ DOE, under contract
DE-AC05-00OR22725 with Oak Ridge National Laboratory, managed and
operated by UT-Battelle, LLC.

\end{acknowledgements}


\end{document}